\documentclass{marine_2025_paper_template}

\usepackage{times}

\usepackage{graphicx} 

\usepackage{array}
\usepackage{booktabs}

\usepackage[round]{natbib}

\usepackage{etoolbox}
\makeatletter
\patchcmd{\@bibitem}{\hskip\labelsep}{}{}{}
\patchcmd{\@biblabel}{\hskip\labelsep}{}{}{}
\makeatother

\usepackage{hyperref}
\hypersetup{
  colorlinks   = true, 
  urlcolor     = black, 
  linkcolor    = black, 
  citecolor   = black 
}

\usepackage{pgfgantt}
\usepackage{rotating}
\usepackage[graphicx]{realboxes}

\newganttchartelement*{mymilestone}{
mymilestone/.style={
shape=isosceles triangle,
inner sep=0pt,
draw=cyan,
top color=white,
bottom color=cyan!50
},
mymilestone incomplete/.style={
/pgfgantt/mymilestone,
draw=yellow,
bottom color=yellow!50
},
mymilestone label font=\slshape,
mymilestone left shift=0pt,
mymilestone right shift=0pt
}

\newgantttimeslotformat{stardate}{%
\def\decomposestardate##1.##2\relax{%
\def\stardateyear{##1}\def\stardateday{##2}%
}%
\decomposestardate#1\relax%
\pgfcalendardatetojulian{\stardateyear-01-01}{#2}%
\advance#2 by-1\relax%
\advance#2 by\stardateday\relax%
}

\usepackage{float}

\usepackage{tikz}
\usepackage{pgfplots}
\pgfplotsset{compat=1.18}

\usepackage{amsmath}

\usepackage{enumitem}

\usepackage{graphicx}
\usepackage{subcaption} 
\usepackage[labelformat=simple,subrefformat=parens]{caption}

\usepackage{bm}

\usepackage{mathtools}
\usepackage{graphicx}

\usepackage{hyperref}

\usepackage{graphicx}
\usepackage{amsmath}
\usepackage{amsfonts}
\usepackage{amssymb}

\usepackage{tikz}
\usepackage{pgfplots}
\pgfplotsset{compat=1.18}

\title{Predicting airfoil pressure distribution using \break boundary graph neural networks}

\author{Sankalp Jena$^{1, *}$, Gabriel D. Weymouth $^{1}$, Artur K. Lidtke $^{2}$, Andrea Coraddu $^1$}

\address{$^1$ Delft University of Technology, The Netherlands, \and $^2$ Maritime Research Institute Netherlands, The Netherlands \and
$^{*}$ s.jena@tudelft.nl}


\abstract{

\noindent  Surrogate models are essential for fast and accurate surface pressure and friction predictions during design optimization of complex lifting surfaces.
This study focuses on predicting pressure distribution over two-dimensional airfoils using graph neural networks (GNNs), leveraging their ability to process non-parametric geometries.
We introduce boundary graph neural networks (B-GNNs) that operate \textit{exclusively} on surface meshes and compare these to previous work on volumetric GNNs operating on volume meshes. All of the training and evaluation is done using the \texttt{airfRANS} (Reynolds-averaged Navier-Stokes) database. 
We demonstrate the importance of all-to-all communication in GNNs to enforce the global incompressible flow constraint and ensure accurate predictions.
We show that supplying the B-GNNs with local physics-based input-features, such as an approximate local Reynolds number $\mathrm{Re}_x$ and the inviscid pressure distribution from a panel method code, enables a $83\%$ reduction of model size and $87\%$ of training set size relative to models using purely geometric inputs to achieve the same in-distribution prediction accuracy.
We investigate the generalization capabilities of the B-GNNs to out-of-distribution predictions on the \texttt{S809/27} wind turbine blade section and find that incorporating inviscid pressure distribution as a feature reduces error by up to $88\%$ relative to purely geometry-based inputs.
Finally, we find that the physics-based model reduces error by $85\%$ compared to the state-of-the-art volumetric model INFINITY.}

\begin{document}

\section{INTRODUCTION}

\noindent Cavitation on ship propellers produces underwater radiated noise that adversely affects marine life \citep{basan2024} and erodes propeller surface \citep{dular2015}. Design exploration and optimization are performed to address the above concerns while striving for high performance. Scale-resolved computational fluid dynamics (CFD) simulations of cavitating propellers are computationally prohibitive for even a modest number of cases, making fast surrogate models an essential tool for design studies. While panel methods like XFOIL \citep{drela1989} can simulate flow over 2D airfoils using only a surface mesh, and similar methods can be applied to 3D lifting surfaces, these tools require empirical corrections for viscous and cavitation effects which are based on small data sets and their 3D generalization error is poorly documented. In this work, we focus on surrogate modelling methodologies with the possibility to scale to complex 3D propeller flow starting with 2D airfoil shapes as a stepping stone on the way to address the full propeller geometry. \newline

\noindent Although Gaussian-process regression is the most popular approach for surrogate modelling, it applies to low-dimensional and fixed parametrization of the geometry. Deep learning approaches are suitable for high-dimensional learning such as NeuralFoil \citep{sharpe2024} which employs multilayer perceptrons (MLPs) to predict pressure distribution over 2D airfoils. While NeuralFoil operates on nonparametric 2D foil mesh, it maps the geometry to a fixed shape parameterization in its pipeline, limiting its ability to make out-of-sample predictions. Alternatively, Graph Neural Networks (GNNs) are specialized neural networks designed to operate on graph-structured data \citep{scarselli2009,bronstein2021}. This makes them well-suited to predicting fluid flows where computational meshes are already in common use, such as \cite{immordino2025} that uses GNNs to predict pressure distributions on a 3D aircraft. While that work and many others actually use geometric parameters as inputs, a fully nonparametric GNN enables predictions on shapes completely outside the sample distribution, as we will show in this paper.\newline 

\noindent Most GNN airfoil applications employ a \textit{volumetric} model; predicting flow quantities, such as velocity and pressure, over the entire numerical domain. Notable examples include the U-Net convolutional neural network (CNN) architecture by \cite{thuerey2020}, a GNN-based approach by \cite{bonnet2023}, and an implicit neural fields model (INFINITY) by \cite{serrano2023} for predicting Reynolds-averaged Navier-Stokes (RANS) flow around 2D airfoils. However, a volumetric approach not only increases the computational cost but also often fails to generalize well to out-of-distribution geometries. Therefore, in the current study, we focus exclusively on the use of a \textit{boundary} model; as used in \cite{durasov2023} who propose a GNN architecture for 2D airfoils inspired by iterative CFD solvers.\newline

\noindent A clear gap in the literature is the limited work that relates the fundamentals of fluid mechanics to the architecture and input set for GNNs.
For example, the governing fluid equations are elliptic due to the incompressibility constraint. This means that every point on the boundary will affect the pressure at every other point; requiring all-to-all communication in the prediction pipeline. The scaling of the network with the size of the boundary graph is therefore a serious concern, and while the iterative model by \cite{durasov2023} proposes one solution, they do not investigate the complexity of the architecture required for accurate pressure predictions and do not consider the scaling of the architecture when extending to handle 3D geometries. In addition, since our target function is physical, the use of physics-based features \citep{weymouth2013} could greatly enhance our data independence and generalization performance. In this work, we address this gap by developing a systematic framework for boundary-graph neural networks (B-GNNs) that demonstrates the importance of all-to-all communication and provides the scaling of the B-GNNs with the size of the geometry. We show that boundary graph U-Nets have optimal all-to-all scaling and physics-based features improve generalization by $7\times$ compared to the state of the art and enable extrapolation to completely out-of-sample geometries.

\section{METHODOLOGY}

\subsection{Problem definition}


The surrogate model operates on a discretized airfoil geometry, which is represented as a mesh containing $N$ nodes. Each node in the mesh has spatial coordinates $(x,y)$ and associated flow information $\mathbf{q}_i$. The input and output features at the $i^{\text{th}}$-node is defined as:

\begin{equation}
    \mathbf{x}_i = [x,y,\mathbf{q}_i]^\mathrm{T} ; \qquad \mathbf{y}_i = [c_p] ,
\end{equation}

\noindent
where $c_p$ is the viscous pressure coefficient and $\mathbf{q}_i$ refers to the flow information that is provided to some of the discussed models. The airfoils are rotated by the freestream angle of attack $(\alpha)$, embedding the information into the geometry. This design allows the model to infer the angle of attack implicitly through the node coordinates. \newline

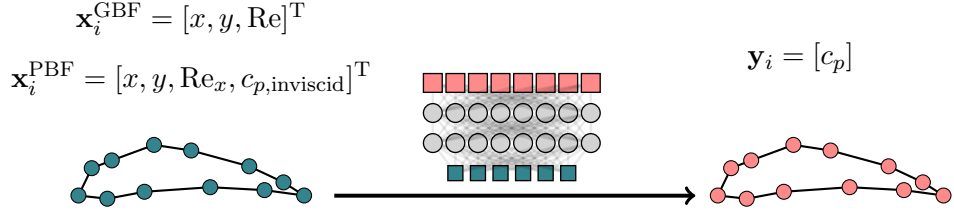
\begin{figure}[t]
    \centering
    \definecolor{myblue}{RGB}{166, 189, 219}
\definecolor{mygray}{RGB}{211, 211, 211}
\definecolor{mypink}{RGB}{255, 166, 201}
\definecolor{myred}{RGB}{255, 138, 138}
\definecolor{steelblue51132141}{RGB}{51,132,141}

\begin{tikzpicture}[scale=1, 
                    nodeout/.style={rectangle, fill=myred, opacity=1, minimum width=2.5mm, minimum height=2.5mm, inner sep=0pt, line width=0.5pt, draw=black}, nodehidden/.style={circle, fill=mygray, opacity=1, minimum size=2.5mm, inner sep=0pt, line width=0.5pt, draw=black}, 
                    nodein/.style={rectangle, fill=steelblue51132141, minimum width=2mm, minimum height=2mm, inner sep=0pt, line width=0.5pt, draw=black}]

    \def\scaleFactor{3} 
    \def\xShift{-1.5}

    \coordinate (P0) at ({\xShift + \scaleFactor * 0.9996}, {\scaleFactor * 0.0008});
    \coordinate (P1) at ({\xShift + \scaleFactor * 0.9083}, {\scaleFactor * 0.0590});
    \coordinate (P2) at ({\xShift + \scaleFactor * 0.7565}, {\scaleFactor * 0.1303});
    \coordinate (P3) at ({\xShift + \scaleFactor * 0.4992}, {\scaleFactor * 0.2004});
    \coordinate (P4) at ({\xShift + \scaleFactor * 0.3345}, {\scaleFactor * 0.2253});
    \coordinate (P5) at ({\xShift + \scaleFactor * 0.1474}, {\scaleFactor * 0.1564});
    \coordinate (P6) at ({\xShift + \scaleFactor * 0.0586}, {\scaleFactor * 0.1220});
    \coordinate (P7) at ({\xShift + \scaleFactor * -0.0004}, {\scaleFactor * 0.0008});
    \coordinate (P8) at ({\xShift + \scaleFactor * 0.1276}, {\scaleFactor * -0.0135});
    \coordinate (P9) at ({\xShift + \scaleFactor * 0.2922}, {\scaleFactor * 0.0174});
    \coordinate (P10) at ({\xShift + \scaleFactor * 0.5837}, {\scaleFactor * 0.0376});
    \coordinate (P11) at ({\xShift + \scaleFactor * 0.8242}, {\scaleFactor * 0.0246});
    \coordinate (P12) at ({\xShift + \scaleFactor * 0.9996}, {\scaleFactor * 0.0008});
    
    \draw[thick] (P0) -- (P1) -- (P2) -- (P3) -- (P4) -- (P5) -- (P6) -- (P7) -- (P8) -- (P9) -- (P10) -- (P11) -- (P12) -- cycle;
    
    \foreach \i in {1, 2, 3, 4, 5, 6, 7, 8, 9, 10, 11, 12} {
        \node[circle, draw, fill=steelblue51132141, inner sep=2pt] at (P\i) {};
    }
    
    \def\xShift{7}

\coordinate (P0) at ({\xShift + \scaleFactor * 0.9996}, {\scaleFactor * 0.0008});
\coordinate (P1) at ({\xShift + \scaleFactor * 0.9083}, {\scaleFactor * 0.0590});
\coordinate (P2) at ({\xShift + \scaleFactor * 0.7565}, {\scaleFactor * 0.1303});
\coordinate (P3) at ({\xShift + \scaleFactor * 0.4992}, {\scaleFactor * 0.2004});
\coordinate (P4) at ({\xShift + \scaleFactor * 0.3345}, {\scaleFactor * 0.2253});
\coordinate (P5) at ({\xShift + \scaleFactor * 0.1474}, {\scaleFactor * 0.1564});
\coordinate (P6) at ({\xShift + \scaleFactor * 0.0586}, {\scaleFactor * 0.1220});
\coordinate (P7) at ({\xShift + \scaleFactor * -0.0004}, {\scaleFactor * 0.0008});
\coordinate (P8) at ({\xShift + \scaleFactor * 0.1276}, {\scaleFactor * -0.0135});
\coordinate (P9) at ({\xShift + \scaleFactor * 0.2922}, {\scaleFactor * 0.0174});
\coordinate (P10) at ({\xShift + \scaleFactor * 0.5837}, {\scaleFactor * 0.0376});
\coordinate (P11) at ({\xShift + \scaleFactor * 0.8242}, {\scaleFactor * 0.0246});
\coordinate (P12) at ({\xShift + \scaleFactor * 0.9996}, {\scaleFactor * 0.0008});
    \draw[thick] (P0) -- (P1) -- (P2) -- (P3) -- (P4) -- (P5) -- (P6) -- (P7) -- (P8) -- (P9) -- (P10) -- (P11) -- (P12) -- cycle;
    
    \foreach \i in {1, 2, 3, 4, 5, 6, 7, 8, 9, 10, 11, 12} {
        \node[circle, draw, fill=myred, inner sep=2pt] at (P\i) {};
    }

    \draw[->, thick, black, line width=1.5pt] ({\scaleFactor * 1.1 - 1.4}, {\scaleFactor * 0.0}) -- ({\xShift - \scaleFactor*0.1 + \scaleFactor * 0.0}, {\scaleFactor * 0.0});

    \def\deltax{0.3};
    \def\deltay{0.4};
    \def\mlpstartx{\scaleFactor * 1.17};
    \def\mlpstarty{\scaleFactor * 0.1};

        \node[nodein, fill=steelblue51132141, opacity=1] (ni1) at (\mlpstartx,\mlpstarty) {};
        \node[nodein, fill=steelblue51132141, opacity=1] (ni2) at (\mlpstartx+\deltax,\mlpstarty) {};
        \node[nodein, fill=steelblue51132141, opacity=1] (ni3) at (\mlpstartx+2*\deltax,\mlpstarty) {};

        \node[nodein, opacity=1] (ni4) at (\mlpstartx+3*\deltax,\mlpstarty) {};
        \node[nodein, opacity=1] (ni5) at (\mlpstartx+4*\deltax,\mlpstarty) {};
        \node[nodein, opacity=1] (ni6) at (\mlpstartx+5*\deltax,\mlpstarty) {};

        \foreach \x in {1,...,8} {
            \pgfmathtruncatemacro{\layer}{1}
            \node[nodehidden] (nh\layer\x) at ({\mlpstartx+(\x-1)*\deltax-\deltax},{\mlpstarty+\layer*\deltay}) {};
            \foreach \prevnode in {1,...,6} {
                \draw[line width=1pt, color=mygray, opacity=0.085, draw=black] (ni\prevnode) -- (nh\layer\x);
            }
        }

        \foreach \x in {1,...,8} {
            \pgfmathtruncatemacro{\layer}{2}
            \node[nodehidden] (nh\layer\x) at ({\mlpstartx+(\x-1)*\deltax-\deltax},{\mlpstarty+\layer*\deltay}) {};
            \pgfmathtruncatemacro{\prevlayer}{\layer-1}
            \foreach \prevnode in {1,...,8} {
                \draw[line width=1pt, color=mygray, opacity=0.085, draw=black] (nh\prevlayer\prevnode) -- (nh\layer\x);
            }
        }

        \foreach \x in {1,...,8} {
            \pgfmathtruncatemacro{\layer}{3}
            \node[nodeout] (nh\layer\x) at ({\mlpstartx+(\x-1)*\deltax-\deltax},{\mlpstarty+\layer*\deltay}) {};
            \pgfmathtruncatemacro{\prevlayer}{\layer-1}
            \foreach \prevnode in {1,...,8} {
                \draw[line width=1pt, color=mygray, opacity=0.085, draw=black] (nh\prevlayer\prevnode) -- (nh\layer\x);
            }
        }

    \coordinate (V_GBF) at ({\scaleFactor * 0.0}, {\scaleFactor * 0.0 + 2});
    \node at (V_GBF) [anchor=south] {$\mathbf{x}^{\mathrm{GBF}}_i = [x,y,\mathrm{Re}]^\mathrm{T}$};

    \coordinate (V_PBF) at ({\scaleFactor * 0.0}, {\scaleFactor * 0.0 + 1.2});
    \node at (V_PBF) [anchor=south] {$\mathbf{x}^{\mathrm{PBF}}_i = [x,y,\mathrm{Re}_x, c_{p, \mathrm{inviscid}}]^\mathrm{T}$};



    \coordinate (VO) at ({\xShift + \scaleFactor * 1.2 - 2.5}, 1.5);
    \node at (VO) [anchor=south] {$\mathbf{y}_i =[c_p]$};

    \end{tikzpicture}
    \vspace{-0.8cm}
    \caption{B-GNN framework used to learn the pressure distribution of an airfoil in steady viscous flow. $i^{\text{th}}$ node with geometry-based feature (GBF), $\mathbf{x}^{\mathrm{GBF}}_i$, containing node coordinates ($x,y$) and Reynolds number ($\mathrm{Re}$), is mapped to node prediction, $\mathbf{y}_i$, with coefficient of pressure, $c_p$. Models with physics-based features (PBFs), $\mathbf{x}^{\mathrm{PBF}}_i$, additionally use local Reynolds number $\mathrm{Re}_x$ and inviscid coefficient of pressure $c_{p,\text{inviscid}}$ as input feature.}
    \label{fig:airfoil_gnn}
\end{figure}

\noindent The airfoil mesh can be represented as a graph, where the nodes correspond to discrete points on the airfoil surface and the edges define connectivity between them. The adjacency matrix $\mathbf{A}$ encodes the graph structure \citep{bronstein2021}. For a two-dimensional airfoil, the connectivity follows a natural order, running from the trailing edge to the leading edge and back, forming a closed-loop structure. As a result, the airfoil mesh is a ring graph \citep{bronstein2021}, as illustrated in Figure \ref{fig:airfoil_gnn}.
The input node embedding matrix \( \mathbf{X} \) contains the node features $\mathbf{x}_i$. A convolution layer of GCN is a function $\mathbf{F}$ with parameters $\mathbf{\Phi}$. It takes node embeddings \( \mathbf{X} \) and the adjacency matrix \( \mathbf{A} \) as the input to generate new node embeddings. The node embedding matrix $\mathbf{H}$ after $K$ convolutions is given by

\begin{align}
  \mathbf{H}_1 & = \mathbf{F}[\mathbf{X}, \mathbf{A}, \mathbf{\Phi}_{0}] \\
               & \shortvdotswithin{ = } \notag \\[-3.5ex]
  \mathbf{H}_K & = \mathbf{F}[\mathbf{H}_{K-1}, \mathbf{A}, \mathbf{\Phi}_{K-1}]
\end{align}

\noindent
Initially, each node's embedding only contains self-information. As the GCN operates, node embeddings are updated by passing messages from neighbouring nodes. This is done using edge convolution \citep{wang2019}. The node embedding at $K$-th convolution is given by

\begin{equation}
    \mathbf{h}_{i,K} = \underset{j:(i,j)\in \mathcal{E}}{\mathrm{max}} \bar{h}_{\Theta}(\mathbf{x}_i, \mathbf{x}_j - \mathbf{x}_i)
\end{equation}

\noindent where $\bar{h}_{\Theta}: \mathbb{R}^{|\mathbf{x}_i|} \times \mathbb{R}^{|\mathbf{x}_j|} \rightarrow \mathbb{R}^{|\mathbf{h}_{i,K}|}$ is an MLP with a set of learnable parameters $\Theta$ and exponential linear unit (ELU) \citep{clevert2016} activation function, and $\mathcal{E}$ is the set of edges. Messages are aggregated using a symmetric operation $\mathrm{max}$. The central node feature $\mathbf{x}_i$ captures the global shape information with respect to the entire graph, and local neighbourhood information with respect to its neighbours accounted by the difference with neighbourhood $\mathbf{x}_j - \mathbf{x}_i$.

\subsection{Dataset}

We utilize the \texttt{airfRANS} dataset \citep{bonnet2023}, which provides steady-state turbulent flow solutions over two-dimensional \texttt{NACA4/5} airfoils computed using Reynolds-averaged Navier–Stokes equations.  To construct airfoil graphs, each airfoil is discretized by random sampling between $120$ and $160$ nodes from the surface, with a higher density of points near the leading and trailing edges to capture steep pressure gradients accurately. Having varying numbers of nodes across airfoil graphs serves as a test of the ability of B-GNNs to deal with inconsistent geometry descriptions in the dataset. The pressure coefficient, $c_p$, is computed by normalizing pressure values with the dynamic pressure, $q_{\infty} = 0.5 \rho U_{\infty}^2$, where $\rho$ is the density of the fluid and $U_{\infty}$ is the freestream velocity. The dataset consists of predefined $800$ training samples and $200$ test samples, with the training set further split into $90\%$ for training and $10\%$ for validation. All features and labels are normalized to the range $[0,1]$ using min-max normalization based on the training set statistics.

\subsection{Input features}

\noindent \textbf{Geometry-based features}

The model arguments are the spatial coordinates $(x,y)$ and the Reynolds number $(\mathrm{Re})$. The geometry-based feature (GBF) is given by
        \begin{equation}
        \mathbf{x}^{\mathrm{GBF}}_i = [x,y,\mathrm{Re}]^\mathrm{T}
        \end{equation}
\newpage
\noindent \textbf{Physics-based features} 

To investigate the influence of \textit{a priori} knowledge of the problem, B-GNNs are supplied with physics-based features (PBFs) \citep{weymouth2013} derived from potential flow solutions obtained using XFOIL \citep{drela1989}. These features include the approximate local Reynolds Number with respect to the stagnation point ($x_\text{stag}$), $\mathrm{Re}_x = (x-x_\text{stag}) U_{\infty}/\nu$, where $\nu$ is the kinematic viscosity of the fluid, and the inviscid pressure coefficient, $c_{p,\text{inviscid}}$. These features encode fundamental physics, such as the growth of the boundary layer and non-viscous pressure distributions, which could enable B-GNNs models to learn viscous pressure distribution. The physics-based feature (PBF) is given by
        \begin{equation}
        \mathbf{x}^{\mathrm{PBF}}_i = [x,y,\mathrm{Re}_x, c_{p, \mathrm{inviscid}}]^\mathrm{T}
        \end{equation}
        
\subsection{Model architectures}

From the global incompressibility constraint, it follows that each node of the discretized airfoil geometry affects all the other nodes. Therefore, we evaluate two architectures that aggregate far-away neighbourhood information with different complexities: single-level boundary graph convolutional neural network (B-GCN) and multi-level boundary graph-U-Net (B-GUN). The level refers to the number of neighbourhood nodes from which messages are aggregated in one convolution layer.\newline 

\begin{figure}[t]
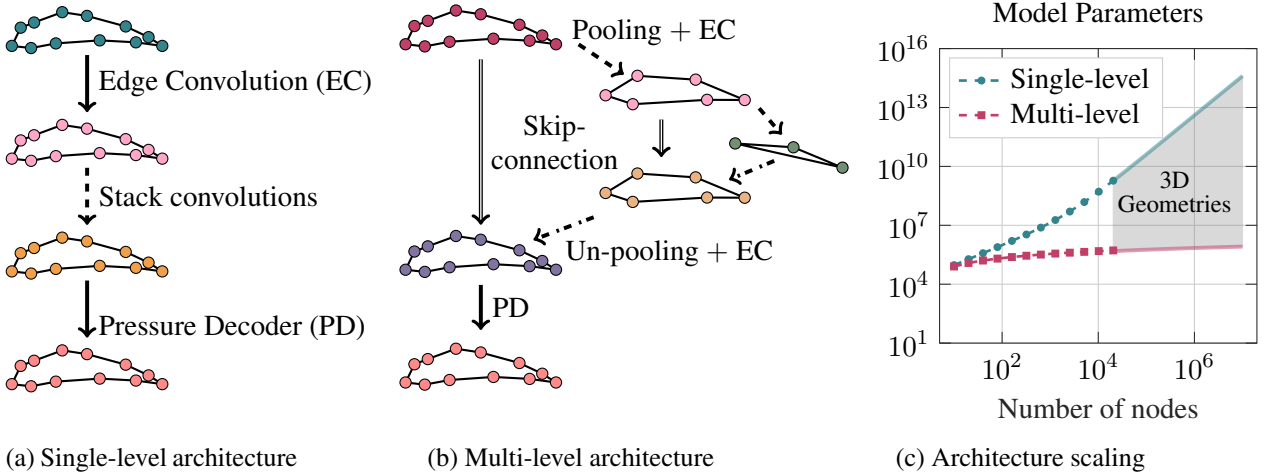

    \centering
    \hspace{-1cm}
    \begin{subfigure}{.228\textwidth}
        \centering
        \definecolor{myblue}{RGB}{166, 189, 219}
\definecolor{mygray}{RGB}{211, 211, 211}
\definecolor{mypink}{RGB}{255, 166, 201}

\definecolor{indianred19364103}{RGB}{193,64,103}
\definecolor{steelblue51132141}{RGB}{51,132,141}
\definecolor{stackconv}{RGB}{240, 160, 75}
\definecolor{myred}{RGB}{255, 138, 138}

\begin{tikzpicture}[scale=1, 
                    nodeout/.style={rectangle, fill=mypink, opacity=1, minimum width=2.5mm, minimum height=2.5mm, inner sep=0pt, line width=0.5pt, draw=black}, nodehidden/.style={circle, fill=mygray, opacity=1, minimum size=2.5mm, inner sep=0pt, line width=0.5pt, draw=black}, 
                    nodein/.style={rectangle, fill=myblue, minimum width=2mm, minimum height=2mm, inner sep=0pt, line width=0.5pt, draw=black}]

    \def\scaleFactor{2} 
    \def\circleDia{1.5}
    \coordinate (P0) at ({\scaleFactor * 0.9996}, {\scaleFactor * 0.0008});
    \coordinate (P1) at ({\scaleFactor * 0.9083}, {\scaleFactor * 0.0590});
    \coordinate (P2) at ({\scaleFactor * 0.7565}, {\scaleFactor * 0.1303});
    \coordinate (P3) at ({\scaleFactor * 0.4992}, {\scaleFactor * 0.2004});
    \coordinate (P4) at ({\scaleFactor * 0.3345}, {\scaleFactor * 0.2253});
    \coordinate (P5) at ({\scaleFactor * 0.1474}, {\scaleFactor * 0.1564});
    \coordinate (P6) at ({\scaleFactor * 0.0586}, {\scaleFactor * 0.1220});
    \coordinate (P7) at ({\scaleFactor * -0.0004}, {\scaleFactor * 0.0008});
    \coordinate (P8) at ({\scaleFactor * 0.1276}, {\scaleFactor * -0.0135});
    \coordinate (P9) at ({\scaleFactor * 0.2922}, {\scaleFactor * 0.0174});
    \coordinate (P10) at ({\scaleFactor * 0.5837}, {\scaleFactor * 0.0376});
    \coordinate (P11) at ({\scaleFactor * 0.8242}, {\scaleFactor * 0.0246});
    \coordinate (P12) at ({\scaleFactor * 0.9996}, {\scaleFactor * 0.0008});
    
    \draw[thick] (P0) -- (P1) -- (P2) -- (P3) -- (P4) -- (P5) -- (P6) -- (P7) -- (P8) -- (P9) -- (P10) -- (P11) -- (P12) -- cycle;
    
    \foreach \i in {1, 2, 3, 4, 5, 6, 7, 8, 9, 10, 11, 12} {
        \node[circle, draw, fill=steelblue51132141, inner sep=\circleDia pt] at (P\i) {};
    }
    
    \def\yShift{-1.5}

    \coordinate (P0) at ({\scaleFactor * 0.9996}, {\yShift + \scaleFactor * 0.0008});
    \coordinate (P1) at ({\scaleFactor * 0.9083}, {\yShift + \scaleFactor * 0.0590});
    \coordinate (P2) at ({\scaleFactor * 0.7565}, {\yShift + \scaleFactor * 0.1303});
    \coordinate (P3) at ({\scaleFactor * 0.4992}, {\yShift + \scaleFactor * 0.2004});
    \coordinate (P4) at ({\scaleFactor * 0.3345}, {\yShift + \scaleFactor * 0.2253});
    \coordinate (P5) at ({\scaleFactor * 0.1474}, {\yShift + \scaleFactor * 0.1564});
    \coordinate (P6) at ({\scaleFactor * 0.0586}, {\yShift + \scaleFactor * 0.1220});
    \coordinate (P7) at ({\scaleFactor * -0.0004}, {\yShift + \scaleFactor * 0.0008});
    \coordinate (P8) at ({\scaleFactor * 0.1276}, {\yShift + \scaleFactor * -0.0135});
    \coordinate (P9) at ({\scaleFactor * 0.2922}, {\yShift + \scaleFactor * 0.0174});
    \coordinate (P10) at ({\scaleFactor * 0.5837}, {\yShift + \scaleFactor * 0.0376});
    \coordinate (P11) at ({\scaleFactor * 0.8242}, {\yShift + \scaleFactor * 0.0246});
    \coordinate (P12) at ({\scaleFactor * 0.9996}, {\yShift + \scaleFactor * 0.0008});
    
    \draw[thick] (P0) -- (P1) -- (P2) -- (P3) -- (P4) -- (P5) -- (P6) -- (P7) -- (P8) -- (P9) -- (P10) -- (P11) -- (P12) -- cycle;
    
    \foreach \i in {1, 2, 3, 4, 5, 6, 7, 8, 9, 10, 11, 12} {
        \node[circle, draw, fill=mypink, inner sep=\circleDia pt] at (P\i) {};
    }

    \def\yShift{-3}

    \coordinate (P0) at ({\scaleFactor * 0.9996}, {\yShift + \scaleFactor * 0.0008});
    \coordinate (P1) at ({\scaleFactor * 0.9083}, {\yShift + \scaleFactor * 0.0590});
    \coordinate (P2) at ({\scaleFactor * 0.7565}, {\yShift + \scaleFactor * 0.1303});
    \coordinate (P3) at ({\scaleFactor * 0.4992}, {\yShift + \scaleFactor * 0.2004});
    \coordinate (P4) at ({\scaleFactor * 0.3345}, {\yShift + \scaleFactor * 0.2253});
    \coordinate (P5) at ({\scaleFactor * 0.1474}, {\yShift + \scaleFactor * 0.1564});
    \coordinate (P6) at ({\scaleFactor * 0.0586}, {\yShift + \scaleFactor * 0.1220});
    \coordinate (P7) at ({\scaleFactor * -0.0004}, {\yShift + \scaleFactor * 0.0008});
    \coordinate (P8) at ({\scaleFactor * 0.1276}, {\yShift + \scaleFactor * -0.0135});
    \coordinate (P9) at ({\scaleFactor * 0.2922}, {\yShift + \scaleFactor * 0.0174});
    \coordinate (P10) at ({\scaleFactor * 0.5837}, {\yShift + \scaleFactor * 0.0376});
    \coordinate (P11) at ({\scaleFactor * 0.8242}, {\yShift + \scaleFactor * 0.0246});
    \coordinate (P12) at ({\scaleFactor * 0.9996}, {\yShift + \scaleFactor * 0.0008});
    
    \draw[thick] (P0) -- (P1) -- (P2) -- (P3) -- (P4) -- (P5) -- (P6) -- (P7) -- (P8) -- (P9) -- (P10) -- (P11) -- (P12) -- cycle;
    
    \foreach \i in {1, 2, 3, 4, 5, 6, 7, 8, 9, 10, 11, 12} {
        \node[circle, draw, fill=stackconv, inner sep=\circleDia pt] at (P\i) {};
    }

    \def\yShift{-4.5}

    \coordinate (P0) at ({\scaleFactor * 0.9996}, {\yShift + \scaleFactor * 0.0008});
    \coordinate (P1) at ({\scaleFactor * 0.9083}, {\yShift + \scaleFactor * 0.0590});
    \coordinate (P2) at ({\scaleFactor * 0.7565}, {\yShift + \scaleFactor * 0.1303});
    \coordinate (P3) at ({\scaleFactor * 0.4992}, {\yShift + \scaleFactor * 0.2004});
    \coordinate (P4) at ({\scaleFactor * 0.3345}, {\yShift + \scaleFactor * 0.2253});
    \coordinate (P5) at ({\scaleFactor * 0.1474}, {\yShift + \scaleFactor * 0.1564});
    \coordinate (P6) at ({\scaleFactor * 0.0586}, {\yShift + \scaleFactor * 0.1220});
    \coordinate (P7) at ({\scaleFactor * -0.0004}, {\yShift + \scaleFactor * 0.0008});
    \coordinate (P8) at ({\scaleFactor * 0.1276}, {\yShift + \scaleFactor * -0.0135});
    \coordinate (P9) at ({\scaleFactor * 0.2922}, {\yShift + \scaleFactor * 0.0174});
    \coordinate (P10) at ({\scaleFactor * 0.5837}, {\yShift + \scaleFactor * 0.0376});
    \coordinate (P11) at ({\scaleFactor * 0.8242}, {\yShift + \scaleFactor * 0.0246});
    \coordinate (P12) at ({\scaleFactor * 0.9996}, {\yShift + \scaleFactor * 0.0008});
    
    \draw[thick] (P0) -- (P1) -- (P2) -- (P3) -- (P4) -- (P5) -- (P6) -- (P7) -- (P8) -- (P9) -- (P10) -- (P11) -- (P12) -- cycle;
    
    \foreach \i in {1, 2, 3, 4, 5, 6, 7, 8, 9, 10, 11, 12} {
        \node[circle, draw, fill=myred, inner sep=\circleDia pt] at (P\i) {};
    }

    \def\yMargin{0.12};
    \draw[->, thick, black, line width=1.5pt]({\scaleFactor * 0.5},0 - \yMargin) -- ({\scaleFactor * 0.5}, -1 + \yMargin) 
    node[midway, right] {Edge Convolution (EC)};

    \def\yShift{-1.5}
    
    \draw[->, dashed, thick, black, line width=1.5pt]({\scaleFactor * 0.5},0 - \yMargin + \yShift) -- ({\scaleFactor * 0.5}, -1 + \yMargin + \yShift)
    node[midway, right] {Stack convolutions};

    \def\yShift{-3}
    
    \draw[->, thick, black, line width=1.5pt]({\scaleFactor * 0.5},0 - \yMargin + \yShift) -- ({\scaleFactor * 0.5}, -1 + \yMargin + \yShift)
    node[midway, right, yshift = -7 pt] {\shortstack{Pressure Decoder (PD)}};

    \end{tikzpicture}
        \caption{Single-level architecture}
        \label{fig:bgcn_architecture}
    \end{subfigure}
    \hspace{0.3cm}
    \begin{subfigure}{.37\textwidth}
        \centering
        \raisebox{0.5cm}{\input{figures/methodology_plots/b-gun-arch}}
        \caption{Multi-level architecture}
        \label{fig:bgun_architecture}
    \end{subfigure}
    \hspace{0.6cm}
    \begin{subfigure}{.24\textwidth}
        \centering
        \raisebox{-0.05cm}{\input{figures/methodology_plots/scaling_vanillaGNN_GUN}}
        \caption{Architecture scaling}
        \label{fig:scaling_bgcn_bgun}
    \end{subfigure}
    \vspace{-0.2cm}
    \caption{ (a) Single-level B-GCN needs recursive convolutions, whereas, (b) multi-level B-GUN coarsens the input graph to capture far-node influence. (c) For 3D geometries, multi-level B-GUN is feasible while single-level B-GCN becomes infeasible.}
    \label{fig:bgnn_architectures}
\end{figure}
    
\noindent {\textbf{Single-level B-GCN}}
    
In a B-GCN (Figure \ref{fig:bgcn_architecture}), a single edge convolution $K$ gathers information from neighbourhood nodes that are 1-hop away. Recursive convolutions are applied to accumulate information from far-away nodes. Node features from every convolution layer are stacked and passed through as pressure decoder MLP to predict the viscous pressure coefficient. The number of convolutions $K$ determines the communication range. In a ring graph with $N$ nodes, full communication requires $K=N/2$ convolutions. Therefore, stacking node embeddings of dimension $H$ from $K$ layers in the pressure decoder MLP leads to quadratic growth in model size (Figure \ref{fig:scaling_bgcn_bgun}), as the MLP requires \{ $\mathcal{O}(K \cdot H)$, $\mathcal{O}(K \cdot H)$, 1\} neurons. While this might not be expensive for 2D geometries, for 3D geometries with hundreds of thousands of nodes, the model would be inefficient. This limitation motivates the need for a more efficient communication mechanism: the multi-level B-GUN. \newline

\newpage
\noindent \textbf{Multi-level B-GUN}
    
In contrast to B-GCN, an edge convolution in B-GUN (Figure \ref{fig:bgun_architecture}) gathers information from nodes more than 1-hop away. This is achieved by coarsening the graph using a binary fusion pooling operation. In binary pooling, two nodes are merged into one, and their features are averaged. This process continues until the graph is reduced to three nodes, as at this level, a single edge convolution aggregates messages from all neighbours. To reconstruct the original graph structure, un-pooling operations are performed from the coarsened graph with three nodes. During un-pooling, the graph is upscaled to match the number of nodes at the corresponding coarse level. The node features are copied from the coarser graph and passed to the finer level with an addition operation, creating skip-connections. After $\log_2(N/3)$ un-pooling operations, the original graph structure is restored. At this point, the node embeddings ($H$) are passed through a pressure decoder MLP with $\{ H, H, 1 \}$ neurons to predict the pressure. Unlike B-GCN, the node embedding size for the pressure decoder MLP in B-GUN does not scale with $N$, making it computationally efficient and suitable for 3D geometries with hundreds of thousands of nodes (Figure \ref{fig:scaling_bgcn_bgun}).

\subsection{Model training details}

Four models are analyzed based on two design choices --- architecture and input features --- \{B-GCN, B-GUN\} $\times$ \{Geometry-based features, Physics-based features\}. The mean squared error [REF] between the predicted and ground truth coefficient of pressure $(c_p)$ at each node is used as the loss function

\begin{equation}
    \mathcal{L} \coloneqq \frac{1}{|\mathcal{B}|} \sum_{i \in \mathcal{B}} {(\mathbf{y}_i - \hat{\mathbf{y}}_i)^2}
\end{equation}

\noindent
where $\mathcal{B}$ is the set of indices of boundary nodes, $\mathbf{y}_i$ is the RANS ground truth and $\hat{\mathbf{y}}$ is the model prediction. Adam optimizer with a learning rate of $10^{-4}$ is employed, alongside a batch size of $32$ for the training. All models are trained for $32000$ epochs. From the model checkpoints, the model with minimum validation loss is used for inference. In both architectures, the key hyperparameters are from the edge convolution function viz. the node embedding size ($H$), edge convolution MLP width ($W$), and edge convolution MLP layers ($L$). The optimal hyperparameter combination obtained at $20$ convolutions is used for the single-level model (B-GCN) and at full depth is used for the multi-level model (B-GUN). The criterion is minimizing validation loss. To reduce computational effort models are trained for $1000$ epochs. \newline

\noindent
\textbf{Sensitivity to training dataset size}

To analyse the influence of physics-based features (PBFs) on the training data requirement, we study the scaling of generalization error with shards of the training set. Following \cite{hestness2017}, we create shards of sizes $\{50, 100, 200, 400\}$ such that training data are added to the smaller shard following a power-law. Models are trained on $5$ randomly sampled shards to provide statistics. A predefined test set of $200$ samples is used for evaluation. \newline

\noindent\textbf{Baselines}

The predictions by four B-GNNs, \{single-level geometry-based (GBF-B-GCN), multi-level geometry-based (GBF-B-GUN), single-level physics-based (PBF-B-GCN), multi-level physics-based (PBF-B-GUN)\}, are compared with two volume-based baselines -- implicit neural fields model INFINITY \citep{serrano2023} and volumetric Graph-U-Net \citep{bonnet2023}.

\section{RESULTS and DISCUSSIONS}

\subsection{All-to-all node communication improves predictions}

\noindent To test the hypothesis that incompressibility constraints require all-to-all node communication for accurate pressure prediction, we vary the parameters that control message passing among nodes in the single-level and multi-level architectures with geometry-based features, GBF-B-GCN and GBF-B-GUN respectively. In the single-level architecture, the pressure decoder MLP scales quadratically with $K$. To balance computational effort and performance, we test $K=\{1, 5, 10, 20\}$ convolutions in this study. Unlike the single-level architecture, the multi-level architecture does not require stacking embeddings, and its pressure decoder MLP remains compact. This allows us to test the model up to full depth $D=\{0,1,2,3,4,5\}$.\newline

\begin{figure}[t]
    \centering
    \begin{subfigure}{0.25\textwidth}
        \centering
        \vspace{1cm}
        \raisebox{0cm}{
\begin{tikzpicture}

\definecolor{cadetblue117169144}{RGB}{117,169,144}
\definecolor{darkslategray38}{RGB}{38,38,38}
\definecolor{darkslategray4584119}{RGB}{45,84,119}
\definecolor{darkslategray66}{RGB}{66,66,66}
\definecolor{lightgray204}{RGB}{204,204,204}
\definecolor{seagreen48110126}{RGB}{48,110,126}
\definecolor{slategray78138134}{RGB}{78,138,134}

\begin{axis}[
width=5cm,
height=5cm,
axis line style={darkslategray38},
log basis y={10},
tick align=inside,
unbounded coords=jump,
x grid style={lightgray204},
xlabel=\textcolor{darkslategray38}{Convolutions (\(\displaystyle K\))},
xmajorticks=true,
xmin=-0.5, xmax=3.5,
xminorgrids,
xmajorgrids,
xtick style={color=darkslategray38},
xtick={0,1,2,3},
xticklabels={1,5,10,20},
y grid style={lightgray204},
ylabel=\textcolor{darkslategray38}{\(\displaystyle \mathcal{L}_{\mathrm{test}}\)},
ymajorticks=true,
ymin=1e-5, ymax=0.003,
yminorgrids,
ymode=log,
ytick style={color=darkslategray38},
]

\def\epsVal{1e-8}

\draw[draw=white,fill=cadetblue117169144] (axis cs:-0.4,\epsVal) rectangle (axis cs:0.4,0.000608911854214966);
\draw[draw=white,fill=slategray78138134] (axis cs:0.6,\epsVal) rectangle (axis cs:1.4,0.000430749409133568);
\draw[draw=white,fill=seagreen48110126] (axis cs:1.6,\epsVal) rectangle (axis cs:2.4,0.000420872267568484);
\draw[draw=white,fill=darkslategray4584119] (axis cs:2.6,\epsVal) rectangle (axis cs:3.4,0.000245962379267439);
\addplot [line width = 1.5pt, dashed, black]
table {%
-0.5 0.00156303563624041
3.5 0.00156303563624041
};
\addplot [line width=1.08pt, darkslategray66]
table {%
0 nan
0 nan
};
\addplot [line width=1.08pt, darkslategray66]
table {%
1 nan
1 nan
};
\addplot [line width=1.08pt, darkslategray66]
table {%
2 nan
2 nan
};
\addplot [line width=1.08pt, darkslategray66]
table {%
3 nan
3 nan
};
\draw (axis cs:1.35,0.0009) node[
  scale=1,
  anchor=west,
  text=darkslategray38,
  rotate=0.0
]{$ \langle c_p \rangle _ {\mathrm{MSE}}$};
\end{axis}

\end{tikzpicture}}
        \caption{Convolution study}
        \label{fig:bgnn_test_loss}
    \end{subfigure}
    \hfill
    \begin{subfigure}{0.25\textwidth}
        \centering
        \vspace{1cm}
        \raisebox{0cm}{
\begin{tikzpicture}

\definecolor{darksalmon217148123}{RGB}{217,148,123}
\definecolor{darkslateblue9951100}{RGB}{99,51,100}
\definecolor{darkslategray38}{RGB}{38,38,38}
\definecolor{darkslategray66}{RGB}{66,66,66}
\definecolor{dimgray13162108}{RGB}{131,62,108}
\definecolor{indianred16174110}{RGB}{161,74,110}
\definecolor{indianred19290108}{RGB}{192,90,108}
\definecolor{indianred209117109}{RGB}{209,117,109}
\definecolor{lightgray204}{RGB}{204,204,204}

\begin{axis}[
width=5cm,
height=5cm,
axis line style={darkslategray38},
log basis y={10},
tick align=inside,
unbounded coords=jump,
x grid style={lightgray204},
xlabel=\textcolor{darkslategray38}{Depth (\(\displaystyle D\))},
xmajorticks=true,
xmin=-0.5, xmax=5.5,
xminorgrids,
xmajorgrids,
xtick style={color=darkslategray38},
xtick={0,1,2,3,4,5},
xticklabels={
  \(\displaystyle {0}\),
  \(\displaystyle {1}\),
  \(\displaystyle {2}\),
  \(\displaystyle {3}\),
  \(\displaystyle {4}\),
  \(\displaystyle {5}\)
},
y grid style={lightgray204},
ylabel=\textcolor{darkslategray38}{\(\displaystyle \mathcal{L}_{\mathrm{test}}\)},
ymajorticks=true,
ymin=1e-5, ymax=0.003,
yminorgrids,
ymode=log,
ytick style={color=darkslategray38},
]

\def\epsVal{1e-8}

\draw[draw=white,fill=darksalmon217148123] (axis cs:-0.4,\epsVal) rectangle (axis cs:0.4,0.000597328355);
\draw[draw=white,fill=indianred209117109] (axis cs:0.6,\epsVal) rectangle (axis cs:1.4,0.000545040821);
\draw[draw=white,fill=indianred19290108] (axis cs:1.6,\epsVal) rectangle (axis cs:2.4,0.000514299027);
\draw[draw=white,fill=indianred16174110] (axis cs:2.6,\epsVal) rectangle (axis cs:3.4,0.000353918178);
\draw[draw=white,fill=dimgray13162108] (axis cs:3.6,\epsVal) rectangle (axis cs:4.4,0.000186683203);
\draw[draw=white,fill=darkslateblue9951100] (axis cs:4.6,\epsVal) rectangle (axis cs:5.4,2.07822668e-05);
\addplot [line width = 1.5pt, dashed, black]
table {%
-0.5 0.0015630356362404
5.5 0.0015630356362404
};
\addplot [line width=1.08pt, darkslategray66]
table {%
0 nan
0 nan
};
\addplot [line width=1.08pt, darkslategray66]
table {%
1 nan
1 nan
};
\addplot [line width=1.08pt, darkslategray66]
table {%
2 nan
2 nan
};
\addplot [line width=1.08pt, darkslategray66]
table {%
3 nan
3 nan
};
\addplot [line width=1.08pt, darkslategray66]
table {%
4 nan
4 nan
};
\addplot [line width=1.08pt, darkslategray66]
table {%
5 nan
5 nan
};

\draw (axis cs:2.5,0.0009) node[
  scale=1,
  anchor=west,
  text=darkslategray38,
  rotate=0.0
]{$ \langle c_p \rangle _ {\mathrm{MSE}}$};
\end{axis}

\end{tikzpicture}}
        \caption{Depth study}
        \label{fig:bgun_test_loss}
    \end{subfigure}\hfill
    \begin{subfigure}{0.38\textwidth}
        \centering
        \vspace{1cm}
        \input{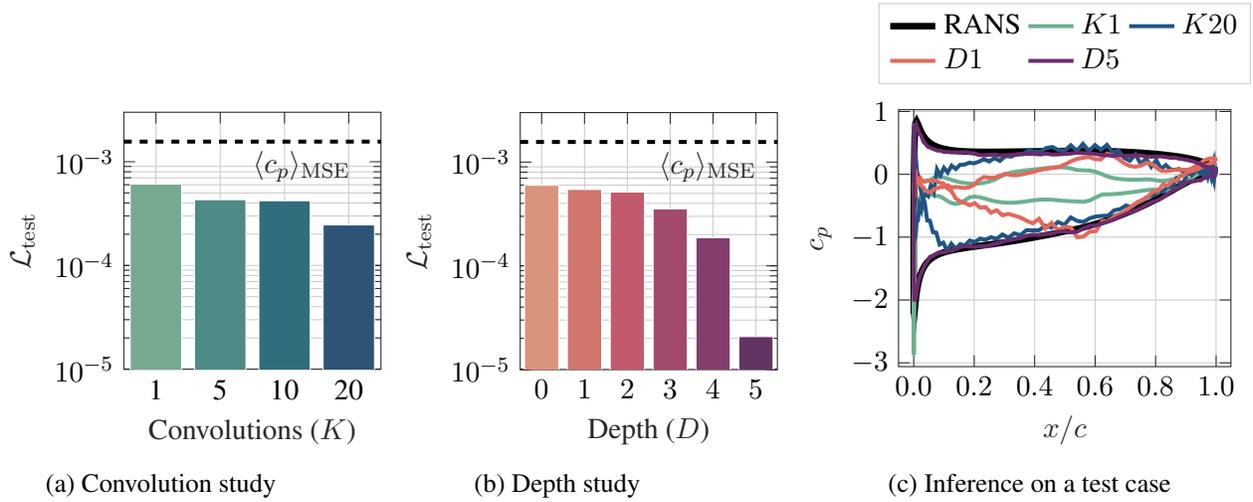}
        \caption{Inference on a test case}
        \label{fig:compare_bgnn_bgun_cp}
    \end{subfigure}
    \vspace{-0.2cm}
    \caption{(a) Test loss decreases with recursive convolutions in the single-level model with the errors being lower than the reference $\langle c_p \rangle_{\text{MSE}}$. (b) Similarly, test loss decreases with depth in the multi-level model. (c) The multi-level model at full depth $D=5$ agrees the most with the ground truth RANS.}
    \label{fig:conv_depth_study}
\end{figure}

Figure \ref{fig:bgnn_test_loss} shows the test loss for the single-level models with increasing convolutions, demonstrating that accuracy improves with greater node communication. The reference error, $\langle c_p \rangle_{\text{MSE}}$, represents the loss if the model simply predicted the mean pressure distribution from the training set at every inference. Since the model achieves lower error, it is not merely regressing to the mean pressure distribution. This is further confirmed in Figure \ref{fig:compare_bgnn_bgun_cp}, which illustrates inference on a test geometry: \texttt{NACA (6.914, 5.245, 8.861)} at $\mathrm{Re}=2.76 \times 10^6$, $\alpha=6 ^{\circ}$. While predictions at $K=20$ are closer to the ground truth, they remain inaccurate and noisy, failing to capture the ground truth pressure distribution. This highlights that without all-to-all communication, the single-level model cannot produce reliable predictions. Similarly, the multi-level geometry-based model achieves higher accuracy as node communication increases with depth, as shown in Figure \ref{fig:bgun_test_loss}. Moreover, the predictions at full depth ($D=5$) closely match the ground truth (Figure \ref{fig:compare_bgnn_bgun_cp}). This improvement is due to all-to-all communication, which enables the model to capture pressure distribution effectively.

\subsection{Physics-based features enable reduction of model size and training data}

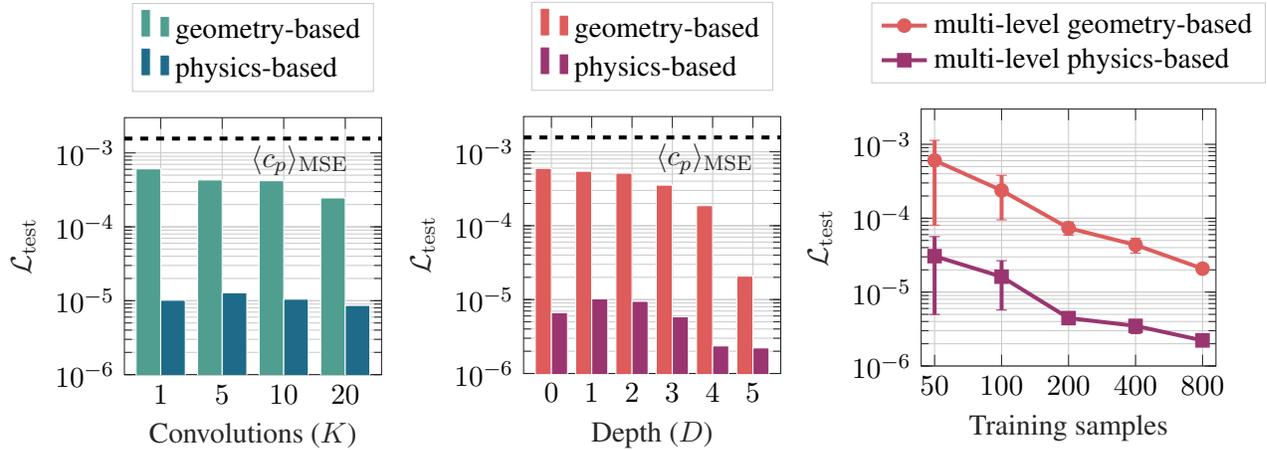
\begin{figure}[t]
    \begin{subfigure}{.3\textwidth}
        \centering
\begin{tikzpicture}

\definecolor{cadetblue79158143}{RGB}{79,158,143}
\definecolor{darkslategray38}{RGB}{38,38,38}
\definecolor{lightgray204}{RGB}{204,204,204}
\definecolor{teal29107137}{RGB}{29,107,137}

\begin{axis}[
width=5cm,
height=5cm,
axis line style={darkslategray38},
legend cell align={left},
legend columns=1,
legend style={
  fill opacity=1,
  draw opacity=1,
  text opacity=1,
  at={(0.5,1.1)},
  anchor=south,
  draw=lightgray204,
},
log basis y={10},
tick align=inside,
x grid style={lightgray204},
xlabel=\textcolor{darkslategray38}{Convolutions (\(\displaystyle K\))},
xmajorgrids,
xmajorticks=true,
xmin=-0.59, xmax=3.59,
xtick style={color=darkslategray38},
xtick={0,1,2,3},
xticklabels={$1$,$5$,$10$,$20$},
y grid style={lightgray204},
ylabel=\textcolor{darkslategray38}{\(\displaystyle \mathcal{L}_{\mathrm{test}}\)},
ymajorgrids,
ymajorticks=true,
yminorgrids,
ymode=log,
ytick style={color=darkslategray38},
ymin=1e-06, ymax=0.003,
    ymode=log,
    ytick style={color=darkslategray38},
]

\def\epsVal{1e-8}

\draw[draw=white,fill=cadetblue79158143] (axis cs:-0.4,\epsVal) rectangle (axis cs:0,0.00060891);

\addlegendimage{ybar,ybar legend,draw=white,fill=cadetblue79158143, legend image post style={xscale=1.2,yscale=1.2}}
\addlegendentry{geometry-based}

\draw[draw=white,fill=cadetblue79158143] (axis cs:0.6,\epsVal) rectangle (axis cs:1,0.00043075);
\draw[draw=white,fill=cadetblue79158143] (axis cs:1.6,\epsVal) rectangle (axis cs:2,0.00042087);
\draw[draw=white,fill=cadetblue79158143] (axis cs:2.6,\epsVal) rectangle (axis cs:3,0.00024596);
\draw[draw=white,fill=teal29107137] (axis cs:-2.77555756156289e-17,\epsVal) rectangle (axis cs:0.4,1.01483165e-05);
\addlegendimage{ybar,ybar legend,draw=white,fill=teal29107137, legend image post style={xscale=1.2,yscale=1.2}}
\addlegendentry{physics-based}

\draw[draw=white,fill=teal29107137] (axis cs:1,\epsVal) rectangle (axis cs:1.4,1.28008924e-05);
\draw[draw=white,fill=teal29107137] (axis cs:2,\epsVal) rectangle (axis cs:2.4,1.04925812e-05);
\draw[draw=white,fill=teal29107137] (axis cs:3,\epsVal) rectangle (axis cs:3.4,8.63228706e-06);

\addplot [line width = 1.5pt, dashed, black]
table {
-0.59 0.00156303563624041
3.59 0.00156303563624041
};

\draw (axis cs:1.3,0.0008) node[
  scale=1,
  anchor=west,
  text=darkslategray38
]{$ \langle c_p \rangle _ {\mathrm{MSE}}$};

\end{axis}

\end{tikzpicture}
        \vspace{-0.5cm}
        \caption{Physics improves accuracy of single-level B-GCN.}
        \label{fig:pbf_bgnn_test_loss}
    \end{subfigure}
    \hfill
    \begin{subfigure}{.3\textwidth}
        \centering
\begin{tikzpicture}

\definecolor{darkslategray38}{RGB}{38,38,38}
\definecolor{indianred2229291}{RGB}{222,92,91}
\definecolor{lightgray204}{RGB}{204,204,204}
\definecolor{mediumvioletred15453111}{RGB}{154,53,111}

\begin{axis}[
width=5cm,
height=5cm,
axis line style={darkslategray38},
legend cell align={left},
legend columns = 1,
legend style={
  fill opacity=1,
  draw opacity=1,
  text opacity=1,
  at={(0.5,1.1)},
  anchor=south,
  draw=lightgray204,
},
log basis y={10},
tick align=inside,
x grid style={lightgray204},
xlabel=\textcolor{darkslategray38}{Depth (\(\displaystyle D\))},
xmajorgrids,
xmajorticks=true,
xmin=-0.69, xmax=5.69,
xtick style={color=darkslategray38},
xtick={0,1,2,3,4,5},
xticklabels={
  \(\displaystyle {0}\),
  \(\displaystyle {1}\),
  \(\displaystyle {2}\),
  \(\displaystyle {3}\),
  \(\displaystyle {4}\),
  \(\displaystyle {5}\)
},
y grid style={lightgray204},
ylabel=\textcolor{darkslategray38}{\(\displaystyle \mathcal{L}_{\mathrm{test}}\)},
ymajorgrids,
ymajorticks=true,
yminorgrids,
ymin=1e-06, ymax=0.003,
    ymode=log,
    ytick style={color=darkslategray38},
]
\def\epsVal{1e-8}

\draw[draw=white,fill=indianred2229291] (axis cs:-0.4,\epsVal) rectangle (axis cs:0,0.000597328355);
\addlegendimage{ybar,ybar legend,draw=white,fill=indianred2229291, legend image post style={xscale=1.2,yscale=1.2}}
\addlegendentry{geometry-based}

\draw[draw=white,fill=indianred2229291] (axis cs:0.6,\epsVal) rectangle (axis cs:1,0.000545040821);
\draw[draw=white,fill=indianred2229291] (axis cs:1.6,\epsVal) rectangle (axis cs:2,0.000514299027);
\draw[draw=white,fill=indianred2229291] (axis cs:2.6,\epsVal) rectangle (axis cs:3,0.000353918178);
\draw[draw=white,fill=indianred2229291] (axis cs:3.6,\epsVal) rectangle (axis cs:4,0.000186683203);
\draw[draw=white,fill=indianred2229291] (axis cs:4.6,\epsVal) rectangle (axis cs:5,2.07822668e-05);
\draw[draw=white,fill=mediumvioletred15453111] (axis cs:-2.77555756156289e-17,\epsVal) rectangle (axis cs:0.4,6.62764569e-06);
\addlegendimage{ybar,ybar legend,draw=white,fill=mediumvioletred15453111, legend image post style={xscale=1.2,yscale=1.2}}
\addlegendentry{physics-based}

\draw[draw=white,fill=mediumvioletred15453111] (axis cs:1,\epsVal) rectangle (axis cs:1.4,1.02298527e-05);
\draw[draw=white,fill=mediumvioletred15453111] (axis cs:2,\epsVal) rectangle (axis cs:2.4,9.47453918e-06);
\draw[draw=white,fill=mediumvioletred15453111] (axis cs:3,\epsVal) rectangle (axis cs:3.4,5.85003636e-06);
\draw[draw=white,fill=mediumvioletred15453111] (axis cs:4,\epsVal) rectangle (axis cs:4.4,2.36108303e-06);
\draw[draw=white,fill=mediumvioletred15453111] (axis cs:5,\epsVal) rectangle (axis cs:5.4,2.21809773e-06);
\addplot [line width = 1.5pt, dashed, black]
table {%
-0.69 0.00156303530093282
5.69 0.00156303530093282
};

\draw (axis cs:2.34,0.0008) node[
  scale=1,
  anchor=west,
  text=darkslategray38
]{$ \langle c_p \rangle _ {\mathrm{MSE}}$};
\end{axis}

\end{tikzpicture}
        \vspace{-0.5cm}
        \caption{Physics improves accuracy of multi-level B-GUN.}
        \label{fig:pbf_bgun_test_loss}
    \end{subfigure}
    \hfill
    \begin{subfigure}{0.37\textwidth}
        \centering
        \raisebox{0.1cm}{
\begin{tikzpicture}

\definecolor{darkslategray38}{RGB}{38,38,38}
\definecolor{indianred2229291}{RGB}{222,92,91}
\definecolor{lightgray204}{RGB}{204,204,204}
\definecolor{mediumvioletred15453111}{RGB}{154,53,111}

\begin{axis}[
width=5.5cm,
height=5cm,
axis line style={darkslategray38},
legend cell align={left},
legend columns = 1,
legend style={
  fill opacity=1,
  draw opacity=1,
  text opacity=1,
  at={(0.5,1.1)},
  anchor=south,
  draw=lightgray204
},
    log basis x={10},
    log basis y={10},
    tick align=inside,
    x grid style={lightgray204},
    xlabel=\textcolor{darkslategray38}{Training samples},
    xmajorgrids,
    xmajorticks=true,
    xminorticks=true,
    xmin=43.5275281648062, xmax=918.958683997629,
    xmode=log,
    xtick style={color=darkslategray38},
    xtick={50,100,200,400,800},
    xticklabels={
      \(\displaystyle {50}\),
      \(\displaystyle {100}\),
      \(\displaystyle {200}\),
      \(\displaystyle {400}\),
      \(\displaystyle {800}\)
    },
    y grid style={lightgray204},
    ylabel=\textcolor{darkslategray38}{\(\displaystyle \mathcal{L}_{\mathrm{test}}\)},
    ymajorgrids,
    yminorgrids,
    ymajorticks=true,
    ymin=1e-06, ymax=0.003,
    ymode=log,
    ytick style={color=darkslategray38},
]

\addplot+[line width = 1.5pt,
  indianred2229291, mark options={indianred2229291, scale=1},
  error bars/.cd, 
    y fixed,
    y dir=both, 
    y explicit,
    error bar style={line width=1.5pt}
] table [x=x, y=y,y error=error, col sep=comma] {
    x,  y,       error
    50, 0.0006057525897631422, 0.0005253575150020925
    100, 0.00023783658252796158, 0.0001428717883669156
    200, 7.355117559200152e-05, 1.4805785250195808e-05
    400, 4.349869341240265e-05, 9.808991600950175e-06
    800, 2.07822668e-05, 0
};
\addlegendentry{multi-level geometry-based}

\addplot+[line width = 1.5pt,
  mediumvioletred15453111, mark options={mediumvioletred15453111, scale=1},
  error bars/.cd, 
    y fixed,
    y dir=both, 
    y explicit,
    error bar style={line width=1.5pt}
] table [x=x, y=y,y error=error, col sep=comma] {
    x,  y,       error
    50, 3.066722019866575e-05, 2.5689228637218805e-05
    100, 1.6163716009032213e-05, 1.0445283178417746e-05
    200, 4.454451618585153e-06, 8.414642202987869e-07
    400, 3.4899384445452596e-06, 7.467996986727589e-07
    800, 2.2180977339303354e-06, 0
};
\addlegendentry{multi-level physics-based}
\end{axis}

\end{tikzpicture}}
        \vspace{-0.5cm}
        \caption{Scaling of generalization error with training dataset size.}
        \label{fig:scaling_generalization}
    \end{subfigure}
    \vspace{-0.2cm}
    \caption{Physics-based features improve accuracy of both (a) single-level B-GCN and (b) multi-level B-GUN and (c) act as supplement data.}
    \label{fig:pbf_improves_accuracy}
\end{figure}

\noindent Providing physics-based features improves accuracy by an order of magnitude, Figure \ref{fig:pbf_improves_accuracy}, with physics-based B-GUN at full depth ($D=5$) achieving the best performance. A closer look reveals that the physics-based models at their lowest model size (of $K=1$ and $D=0$ for B-GCN and B-GUN respectively) outperform the geometry-based counterparts at their highest. Thus, physics-based features reduce model size by $83\%$ relative to models using purely geometric inputs. Figure \ref{fig:scaling_generalization} shows that with added training samples the generalization error improves. At every shard of full training samples including the full training set, the error by physics-based model is an order lower. This means about $87\%$ fewer training samples are needed to get to a similar loss by the geometry-based model. Moreover, physics-based features serve as supplement datasets as they shift the requirement of having more samples to improve generalization in the case of the model without physics-based features.

\subsection{Physics-based features are essential for meaningful extrapolation}

\begin{figure}[t]
    \centering
    \hspace{-0.25cm}
    \begin{subfigure}{.23\textwidth}
        \centering
        \raisebox{0.3cm}{\input{figures/results_plots/naca-s8xx-profiles}}
        \vspace{-0.1cm}
        \hspace{-0.5cm}
        \begin{subfigure}{1.8\textwidth}
            \centering
            \vspace{-0.3cm}
\begin{tikzpicture}

\definecolor{darkslategray38}{RGB}{38,38,38}
\definecolor{lightgray204}{RGB}{204,204,204}

\begin{axis}[
width=4cm,
height=4cm,
axis line style={darkslategray38},
colorbar,
colorbar style={
        ytick={0,5,10},
        yticklabels={
          \(\displaystyle {0}\),
          \(\displaystyle {5}\),
          \(\displaystyle {10}\)
        },
        ylabel={$d_{\mathrm{H}} \cdot 10^{2}$},
        y label style={yshift=0.3cm} 
    }, 
colormap={mymap}{[1pt]
  rgb(0pt)=(0.0196078431372549,0.188235294117647,0.380392156862745);
  rgb(1pt)=(0.129411764705882,0.4,0.674509803921569);
  rgb(2pt)=(0.262745098039216,0.576470588235294,0.764705882352941);
  rgb(3pt)=(0.572549019607843,0.772549019607843,0.870588235294118);
  rgb(4pt)=(0.819607843137255,0.898039215686275,0.941176470588235);
  rgb(5pt)=(0.968627450980392,0.968627450980392,0.968627450980392);
  rgb(6pt)=(0.992156862745098,0.858823529411765,0.780392156862745);
  rgb(7pt)=(0.956862745098039,0.647058823529412,0.509803921568627);
  rgb(8pt)=(0.83921568627451,0.376470588235294,0.301960784313725);
  rgb(9pt)=(0.698039215686274,0.0941176470588235,0.168627450980392);
  rgb(10pt)=(0.403921568627451,0,0.12156862745098)
},
point meta max=10.2481172103722,
point meta min=0,
tick align=inside,
x grid style={lightgray204},
xmajorticks=true,
xmin=0, xmax=4,
xtick style={color=darkslategray38},
xtick={0.5,1.5,2.5,3.5},
xticklabel style={rotate=90.0},
xticklabels={\texttt{S1},\texttt{S2},\texttt{N1},\texttt{N2}},
y dir=reverse,
y grid style={lightgray204},
ymajorticks=true,
ymin=0, ymax=4,
ytick style={color=darkslategray38},
ytick={0.5,1.5,2.5,3.5},
yticklabel style={rotate=90.0},
yticklabels={\texttt{S1},\texttt{S2},\texttt{N1},\texttt{N2}}
]
\addplot graphics [includegraphics cmd=\pgfimage,xmin=0, xmax=4, ymin=4, ymax=0] {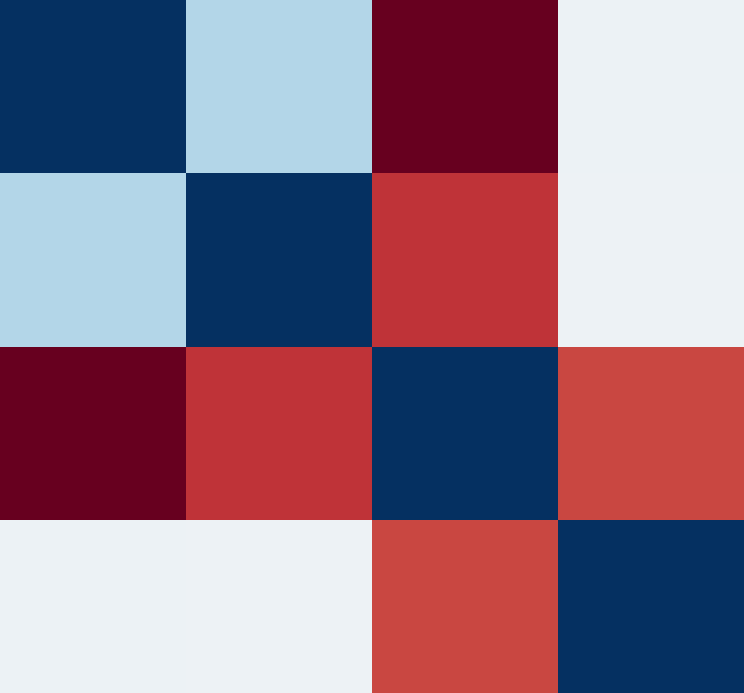};

\def\mScale{0.8}

\draw (axis cs:0.5,0.5) node[
  scale=\mScale,
  text=white,
  rotate=0.0
]{0};
\draw (axis cs:1.5,0.5) node[
  scale=\mScale,
  text=darkslategray38,
  rotate=0.0
]{3.6};
\draw (axis cs:2.5,0.5) node[
  scale=\mScale,
  text=white,
  rotate=0.0
]{10};
\draw (axis cs:3.5,0.5) node[
  scale=\mScale,
  text=darkslategray38,
  rotate=0.0
]{4.8};
\draw (axis cs:0.5,1.5) node[
  scale=\mScale,
  text=darkslategray38,
  rotate=0.0
]{3.6};
\draw (axis cs:1.5,1.5) node[
  scale=\mScale,
  text=white,
  rotate=0.0
]{0};
\draw (axis cs:2.5,1.5) node[
  scale=\mScale,
  text=white,
  rotate=0.0
]{8.8};
\draw (axis cs:3.5,1.5) node[
  scale=\mScale,
  text=darkslategray38,
  rotate=0.0
]{4.9};
\draw (axis cs:0.5,2.5) node[
  scale=\mScale,
  text=white,
  rotate=0.0
]{10};
\draw (axis cs:1.5,2.5) node[
  scale=\mScale,
  text=white,
  rotate=0.0
]{8.8};
\draw (axis cs:2.5,2.5) node[
  scale=\mScale,
  text=white,
  rotate=0.0
]{0};
\draw (axis cs:3.5,2.5) node[
  scale=\mScale,
  text=white,
  rotate=0.0
]{8.6};
\draw (axis cs:0.5,3.5) node[
  scale=\mScale,
  text=darkslategray38,
  rotate=0.0
]{4.8};
\draw (axis cs:1.5,3.5) node[
  scale=\mScale,
  text=darkslategray38,
  rotate=0.0
]{4.9};
\draw (axis cs:2.5,3.5) node[
  scale=\mScale,
  text=white,
  rotate=0.0
]{8.6};
\draw (axis cs:3.5,3.5) node[
  scale=\mScale,
  text=white,
  rotate=0.0
]{0};
\end{axis}

\end{tikzpicture}
            \label{fig:hausdorff_matrix}
        \end{subfigure}
        \caption{\texttt{NACA} v/s \texttt{S8} airfoils.}
        \label{fig:extrapolation_geometries_distance}
    \end{subfigure}
    \hspace{1.35cm}
    \begin{subfigure}{.35\textwidth}
        \centering
        \raisebox{0cm}{\input{figures/results_plots/extrapolation_cp_S809_12}}
        \caption{\texttt{S809} {\texttt{(S1)}} at $\alpha=12^{\circ}$}
        \label{fig:s809_extrapolation}
    \end{subfigure}
    \hspace{-0.45cm}
    \begin{subfigure}{0.35\textwidth}
            \centering
            \raisebox{0.0cm}{\input{figures/results_plots/extrapolation_cp_S827_4}}
            \caption{\texttt{S827} {\texttt{(S2)}} at $\alpha=4^{\circ}$}
            \label{fig:s827_extrapolation}
        \end{subfigure}
    \vspace{-0.2cm}
    \caption{ (a) Hausdorff distance, $d_{\mathrm{H}}$, quantifies the difference between the profiles: \texttt{S809} {\texttt{(S1)}}, \texttt{S827} {\texttt{(S2)}}, \texttt{NACA (1.541, 6.943, 0.0, 5.203)} {\texttt{(N1)}}, \texttt{NACA (3.475, 3.252, 0.0, 19.801)} {\texttt{(N2)}}. (b) and (c) Physics-based features help the multi-level model to agree with the experiment better than the corresponding geometry-based model.}
    \label{fig:pbf_helps_in_extrapolation}
\end{figure}

\noindent To analyze whether enforcing all-to-all communication in B-GNNs compels them to emulate potential flow solvers, we evaluate inference on out-of-distribution geometries commonly used in wind turbine blades: \texttt{S809} \citep{somers1997} and \texttt{S827} \citep{somers2005}. Hausdorff distance \citep{rote1991}, defined as the maximum distance between a point in one set and the nearest point in the other, is used to quantify geometric differences between the \texttt{S809/27} airfoils and the \texttt{NACA4/5} airfoils in the training set. While airfoil thickness provides a general measure of shape variation, it does not fully capture geometric differences. For instance, both \texttt{N2} and \texttt{S809/27} airfoils have similar maximum thicknesses—20\% and 21\% of chord length, respectively—but differ topologically. This distinction is quantified by the Hausdorff distance in Figure \ref{fig:extrapolation_geometries_distance} in which \texttt{S809/27} airfoils are geometrically closer to each other than to the \texttt{NACA} airfoils. Figures \ref{fig:s809_extrapolation} and \ref{fig:s827_extrapolation} compare prediction by full depth geometry-based and physics-based multi-level models with the experiment for two cases: \texttt{S809} at $\mathrm{Re}=2 \times 10^6$ and $\alpha = 12^{\circ}$ \citep{somers1997} and \texttt{S827} at $\mathrm{Re}=3 \times 10^6$ and $\alpha = 4^{\circ}$ \citep{somers2005}. At a high angle of attack $\alpha=12^{\circ}$, the pressure distributions of \texttt{NACA} and \texttt{S809/27} airfoils are similar, making the prediction task less challenging. In this case, both models capture the overall pressure distribution trend; however, the geometry-based model struggles to accurately predict the pressure drop at the leading edge, while the physics-based model aligns much more closely with the experimental data, particularly downstream on the suction surface. Conversely, at a lower angle of attack $\alpha=4^{\circ}$, where the pressure distributions of \texttt{NACA} and \texttt{S809/27} airfoils differ significantly, the extrapolation task becomes more difficult. Here, the geometry-based model overpredicts the pressure drop at the leading edge, leading to inaccurate results. In contrast, the physics-based model uses the inviscid pressure information from the Panel Method to predict a reasonable estimate of the viscous pressure distribution. Note that the Panel Method performs worse than the physics-based model in both cases. These results confirm that all-to-all-communication alone is insufficient for extrapolation -- it requires the help of physics-based features. Given the relatively low computational overhead of obtaining physics-based features and substantial accuracy improvements, incorporating it into surrogate models is an effective strategy to ensure reliable physical predictions in out-of-distribution scenarios.

\subsection{Boundary-GNN improves on baseline model performance}

The baselines predict pressure, $p$, instead of the coefficient of pressure, $c_p$. To enable a direct comparison, the predictions by the geometry-based and physics-based multi-level models are converted to their respective pressure values, $p$, and are then normalized using the mean and standard deviation of the training set before comparison. Additionally, the coefficient of lift, $C_l$, is evaluated from the physical pressures.

\begin{table}[b]
\centering
\caption{Comparison of prediction error (MSE) by the geometry-based and physics-based multi-level models with that of INFINITY \citep{serrano2023} and volumetric Graph-U-Net \citep{bonnet2023}.}
\label{tab:baseline_comparison}
\begin{tabular}{| >{\raggedright\arraybackslash}p{0.7cm} | >{\raggedright\arraybackslash}p{4.3cm} | >{\raggedright\arraybackslash}p{4.3cm} | >{\raggedright\arraybackslash}p{2.7cm} | >{\raggedright\arraybackslash}p{2.7cm} |}
\hline

& {Multi-level physics-based} & {Multi-level geometry-based} & {INFINITY} & {Graph-U-Net} \\

\hline

$p|_{\mathcal{B}}$ & $\mathbf{0.001} \pm \mathbf{0.0001} $ &  $0.020 \pm 0.0043 $ & $0.007 \pm 0.001$ & $0.039 \pm 0.007$ \\ \hline

$c_p$ & $\mathbf{0.001} \pm \mathbf{0.0002} $ & $0.020 \pm 0.0043$ & - & -\\ \hline

$C_l$ & $\mathbf{0.017} \pm \mathbf{0.003} $ & $0.194 \pm 0.047$ & $0.081 \pm 0.007$ & $0.489 \pm 0.105 $\\ \hline

\hline
\end{tabular}
\end{table}

Table \ref{tab:baseline_comparison} highlights the superior performance of the physics-based model over other models by having the lowest MSE for pressure and coefficient of lift. It has $94\%$, $85\%$, and $97\%$ lower $(p|_{\mathcal{B}})$ error compared to the geometry-based model, INFINITY, and volumetric Graph-U-Net respectively. Unlike the geometry-based model, the volumetric models: INFINITY and volumetric Graph-U-Net use surface normals as additional input features, along with spatial coordinates and flow information. This allows them to differentiate between the upper and lower surfaces of the airfoil, which could explain INFINITY’s improved performance over the geometry-based model. Further investigation is needed to determine whether incorporating surface normals into the geometry-based model could enhance its accuracy.


\section{CONCLUSIONS}

This study demonstrates that all-to-all node communication is essential for B-GNNs to accurately learn the mapping of airfoil geometries to pressure distributions. Incorporating physics-based features significantly enhances B-GUN performance by $83\%$. More importantly, they improve generalization to out-of-distribution geometries by up to $88\%$ over the geometry-based model, addressing a major challenge in surrogate modelling. However, reliance on potential flow input-features may bias solutions away from the target flows when they feature strong non-linear effects such as massive flow separation. Additionally, B-GNNs can produce noisy pressure distributions, which could be problematic for future applications such as cavitation modelling. Future work could focus on integrating additional physical features, improving noisy predictions, and extending these methods to three-dimensional unstructured propeller grids.\newline

\section*{ACKNOWLEDGEMENTS}

This research is supported by the granting authority European Research Executive Agency (REA) under the Marie Skłodowska-Curie Actions funded project, SCALE (HORIZON-MSCA-2022 DN-01-0). 


\section*{APPENDIX}

\subsection*{A.1 Model hyperparameter tuning}

The optimal hyperparameters of each B-GNN model are found in the parameter space mentioned in Table~\ref{tab:optimal_hyperparamters}.

\noindent
\begin{minipage}{0.52\textwidth}
\centering

    \begin{table}[H]
    \centering
    \caption{Hyperparameter design space used for the Edge Convolution MLP.}
    \label{tab:optimal_hyperparamters}
    \begin{tabular}{|>{\raggedright\arraybackslash}p{4.6cm}|>{\raggedright\arraybackslash}p{3.0cm}|}
    \hline
    
    \textbf{Hyperparameter} & \textbf{Values} \\
    \hline
    
    Node Embeddings $H$ & ${\{ 4, 8 \}}$ \\ \hline
    Edge conv. MLP Width $W$ & ${\{8, 16, 32, 64, 128 \}}$ \\ \hline
    Edge conv. MLP Layers $L$ & ${\{1, 2 \}}$ \\ \hline
    
    \hline
    \end{tabular}
    \end{table}

\end{minipage}
\hfill
\begin{minipage}{0.42\textwidth}
    
    \begin{table}[H]
    \centering
    \caption{Edge Convolution MLP hyperparameters for different B-GNNs.}
    \label{tab:four_bgnns}
    \begin{tabular}{|>{\raggedright\arraybackslash}p{2.3cm}|>{\raggedright\arraybackslash}p{3.5cm}|}
    \hline
    
    \textbf{Model} & \textbf{Hyperparameters} \\
    \hline
    
    GBF-B-GNN & $H=4$, $W=128$, $L=2$\\ \hline
    GBF-B-GUN & $H=8$, $W=128$, $L=2$\\ \hline
    PBF-B-GNN & $H=4$, $W=128$, $L=1$\\ \hline
    PBF-B-GUN & $H=8$, $W=128$, $L=2$\\ \hline
    
    \hline
    \end{tabular}
    \end{table}
\end{minipage}

\renewcommand{\bibname}{References}
\setlength\bibhang{0pt} 
\bibliographystyle{abbrvnat}
\setcitestyle{author-year}
\bibliography{references}

\end{document}